\documentclass[aps,pra,reprint,twocolumn,superscriptaddress,showpacs,showkeys,longbibliography]{revtex4-1}

\usepackage{amsmath,amssymb,amstext}
\usepackage[usenames,dvipsnames]{color}
\usepackage{graphicx}
\usepackage{bm,bbold,braket}
\usepackage{natbib}
\usepackage{txfonts, comment,stmaryrd}
\usepackage{dcolumn}

\usepackage[english]{babel}

\usepackage[colorlinks,bookmarks=false,citecolor=blue,linkcolor=red,urlcolor=blue]{hyperref}

\begin{document}

\title{Two dimensional bright solitons in dipolar Bose-Einstein condensates with tilted dipoles}

\author{Meghana Raghunandan }
        \affiliation{Indian Institute of Science Education and Research, Pune 411 008, India}        
\author{Chinmayee Mishra}
        \affiliation{Indian Institute of Science Education and Research, Pune 411 008, India}        
\author{Kazimierz \L akomy }
        \affiliation{Institut f\"ur Theoretische Physik, Leibniz Universit\"at Hannover, Appelstrasse 2, DE-30167 Hannover, Germany}        
\author{Paolo Pedri}
        \affiliation{Universit\'e Paris 13, Sorbonne Paris Cit\'e, Laboratoire de Physique des Lasers, F-93430 Villetaneuse, France}
        \affiliation{CNRS, UMR 7538, LPL, F-93430 Villetaneuse, France}
\author{Luis Santos }
     \affiliation{Institut f\"ur Theoretische Physik, Leibniz Universit\"at Hannover, Appelstrasse 2, DE-30167 Hannover, Germany}        
\author{Rejish Nath}
        \affiliation{Indian Institute of Science Education and Research, Pune 411 008, India}
\date{\today}

\begin{abstract}
The effect of dipolar orientation with respect to the soliton plane on the physics of two-dimensional bright solitons in dipolar Bose-Einstein condensates is discussed. Previous studies on such a soliton involved dipoles either perpendicular  or parallel to the condensate-plane. The tilting angle constitutes an additional tuning parameter, which help us to control the in-plane anisotropy of the soliton as well as provides access to previously disregarded regimes of interaction parameters for soliton stability. In addition, it can be used to drive the condensate into phonon instability without changing its interaction parameters or trap geometry.  The phonon-instability in a homogeneous 2D condensate of tilted dipoles always features a transient stripe pattern, which eventually breaks into a metastable soliton gas. Finally, we demonstrate how a dipolar BEC in a shallow trap can eventually be turned into a self-trapped matter wave by an adiabatic approach, involving the tuning of tilting angle.

\end{abstract}



\maketitle

%%%%%%%%%%%%
%% INTRO 
%%%%%%%%%%%%

\section{Introduction}
%bright solitons
Last two decades have witnessed intensive investigations, both theoretical and experimental, on multi-dimensional solitons, especially in systems possessing {\em nonlocal nonlinearity} (NLNL) \cite{nlnl-sol97,bang_nlnl02,nl-nl_15} such as Bose-Einstein condensates (BECs) of particles with permanent \cite{sol-luis,sol-aniso} or induced dipole moments \cite{pohl3d_11}~({\em dipolar BECs}), photo refractive materials \cite{pr_sol_97}, nematic liquid crystals \cite{lc_sol_03,nem_sol12} and others \cite{nl_sol_10, nl_sol_12, nl_sol_13}.  Two-dimensional~(2D) optical bright solitons \cite{os_2d_04} and 
three-dimensional~(3D) light bullets \cite{3d_os_10} have been reported in non-local media.  In addition, nonlocal interactions significantly influence inter-soliton collisions \cite{sol-luis,pr_sol_97,sc_nath_07,coll_ani_12} and may lead to the formation of soliton complexes \cite{fil_kaz_12,cs_skup_07}.

Bright solitons in BECs have been realized experimentally only in quasi-1D alkali-atom gases \cite{sol_expt_02,sol_expt_Nat_02,sol_expt_06}, in which short-range contact-like interactions result in a cubic nonlinearity equivalent to self-focusing nonlinearity in a Kerr media. Whereas in these media higher-dimensional solitons are unstable, 2D solitons may become stable in dipolar BECs for 
a sufficiently large dipole-dipole interaction (DDI) \cite{sol-luis,sol-aniso}. Thus, the realization of dipolar BECs of  chromium (Cr) \cite{crbec_05,crbec_08}, dysprosium (Dy) \cite{dybec_11} and erbium (Er) \cite{erbec_12} opens fascinating perspectives for the realization of 2D BEC solitons. The realization of 2D solitons remains however an open challenge. On one hand the observation of the 2D solitons (for a dipolar orientation 
perpendicular to the soliton plane) proposed in 
Ref.~\cite{sol-luis} demands the inversion of the sign of the DDI by means of rotating fields \cite{ddi_tun02}. On the other hand, anisotropic solitons, proposed in Ref. \cite{sol-aniso} for a dipolar orientation 
on the soliton plane, demand no external driving, 
but additional challenges may arise from their high anisotropy \cite{ani_boris_12}, in particular their vulnerability to collapse instability even in a strictly 2D scenario \cite{nath-pho}. 

\begin{figure}
\vspace{0.5cm}
\centering
\includegraphics[width= 1.\columnwidth]{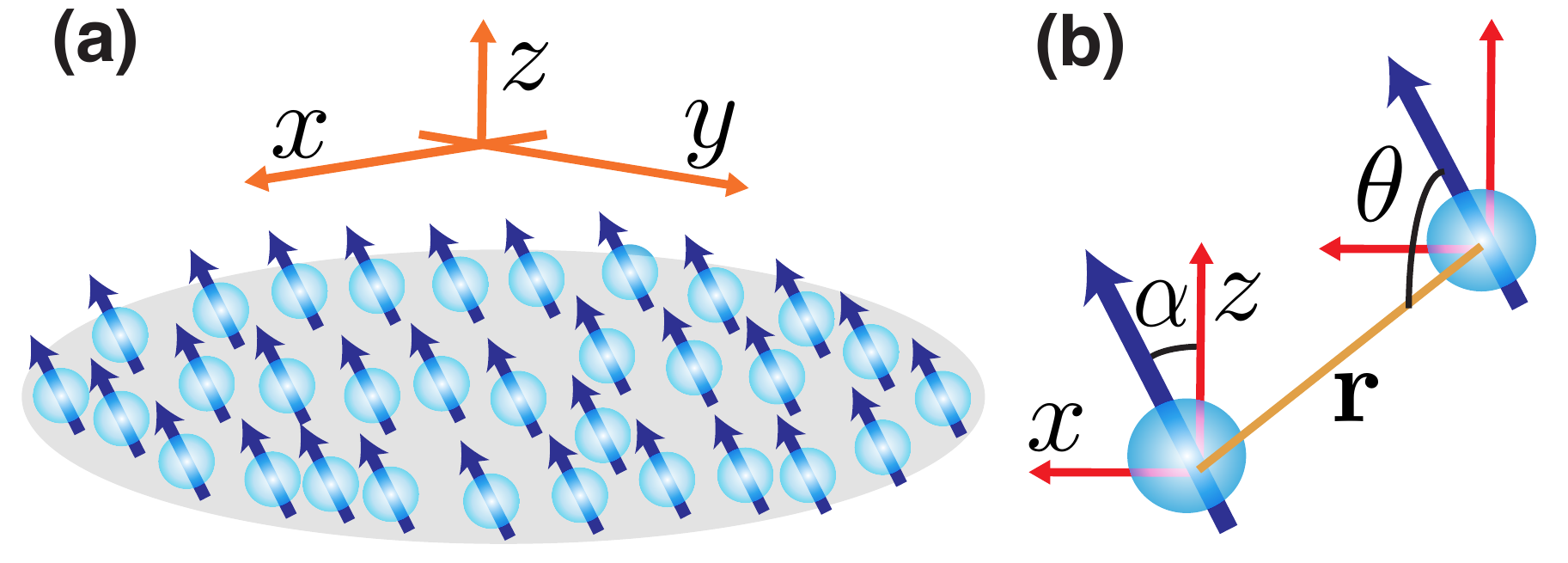}
\caption{\small{(a) The schematic setup of dipolar BEC confined in the $xy$ plane with a strong harmonic confinement along the $z$ axis.  (b) The dipoles are polarized in the $xz$ plane with tilting angle $\alpha$ w.r.t. $z$ axis and $\theta$ is the angle between the dipole vector and the radial vector ${\bf r}$.}}
\label{fig:dset} 
\end{figure}

 In this paper we study the effect of dipolar orientation with respect to the soliton plane  (see Fig. \ref{fig:dset}) on the properties and stability of 2D bright solitons in dipolar BECs. Introducing the tilting angle $\alpha$ with respect to the normal vector of the quasi-2D trap plane \cite{td_fed_14,fd_mac_14,fd_bai_15} allows for accessing previously disregarded regimes of interaction parameters for soliton stability as well as to manipulate the soliton anisotropy in a more controlled manner. These features may enhance the possibilities of realizing 2D BEC solitons experimentally in the state-of-the-art dipolar BECs.

%%%
%%%%

Solitons may be created by driving a BEC into phonon-instability (PI). This has been done in non-polar 1D BECs by tuning the short-range interactions from repulsive to attractive \cite{sol_expt_06}. In contrast, in 2D and 3D PI leads to collapse in non-dipolar BECs \cite{coll_corn_01,coll_hul_00}. On the contrary, the post PI dynamics in 2D dipolar BECs is characterized by the formation of a transient gas of bright solitons \cite{nath-pho}. The resulting solitons exhibit an intriguing dynamics that crucially depends on the nature of DDI on the soliton plane. Isotropic solitons attract each other in any direction, undergoing fusion during collisions and eventually becoming a single large soliton. The scenario is different for anisotropic solitons, as the interaction between the solitons is anisotropic in the plane, which makes them fuse only when they are colliding along the dipolar axis. Here we extend these studies to the tilted case. As we show, the post-PI dynamics is always characterized by a transient stripe pattern that eventually breaks into bright solitons.  

The paper is structured as follows. In Sec. \ref{model} we discuss the model and the corresponding non-local Gross-Pitaevskii equations (NLGPEs). In Sec. \ref{be-pi}, we examine the Bogoliubov excitations of a 2D homogeneous BEC, identifying the regimes of PI as a function of the interaction parameters and the tilting angle. Sec. \ref{2dsol} is devoted to analyzing the stability and the properties of the soliton. The  stable/unstable regions of Cr, Er and Dy BECs are discussed as a function of system parameters. In Sec. \ref{solcr} we demonstrate how to prepare a 2D soliton by varying the tilting angle from a soliton unstable to a soliton stable region in the case of a Cr BEC. Finally we conclude in Sec. \ref{con}.

%%%%%%%%%%%

\section{Model}
\label{model}

We consider a  BEC of  $N$ particles with magnetic or electric dipole moment $d$, oriented in the $xz$ plane forming an angle $\alpha$ with the $z$ axis, using a sufficiently large external field (Fig. \ref{fig:dset}). The DDI potential is $V_d({\bf r})=g_d(1-3\cos^2\theta)/r^3$, where $\theta$ is the angle formed by the dipole vector ${\bf d}\equiv d(\sin\alpha \  \hat x + \cos\alpha \ \hat z)$ and the radial vector {\bf r}, with $g_d\propto Nd^2$ being the strength of the dipole-potential. At low-enough temperatures the system is described by a non-local Gross-Pitaevskii equation (NLGPE):
\begin{equation}
\begin{split}
i\hbar\frac{\partial}{\partial t}\Psi({\bf r},t)=\left[-\frac{\hbar^2}{2m}\nabla^2+V_{t}({\bf r})+g|\Psi({\bf r},t)|^2+\right .\\
\left. \int d{\bf r}^{\prime}V_d({\bf r}-{\bf r}^{\prime})|\Psi({\bf r}^{\prime},t)|^2\right]\Psi({\bf r},t),
\label{gpe3d}
\end{split}
\end{equation}
where $\int d{\bf r}|\Psi({\bf r},t)|^2=1$ and $g=4\pi\hbar^2aN/m$ is the coupling constant that characterizes the short-range contact interaction, with $a$ the $s$-wave scattering length. Further, we assume a strong harmonic confinement along the $z$ direction with a frequency $\omega_z$ and no trapping in the $xy$ plane, hence $V_t(r)=m\omega_zz^2/2$. This trapping is sufficiently strong 
such that the system remains in the ground state, 
$\phi_0(z)=\exp(-z^2/l_z^2)/\sqrt{\pi^{1/2}l_z}$, of the harmonic oscillator along the $z$-axis with $l_z=\sqrt{\hbar/m\omega_z}$, factorizing the BEC wave function as $\Psi({\bf r})=\psi(x,y)\phi_0(z)$. In that case, the physics of dipolar BEC becomes quasi-2D, with the restriction that $|\mu_{2D}|\ll\hbar\omega_z$, where $\mu_{2D}$ is the chemical potential of the 2D gas. 
Employing this factorization, convolution theorem, the Fourier transform of the DDI potential, 
 \begin{equation}
 \tilde V_d({\bf k})=\frac{4\pi g_d}{3}\left[\frac{3\left(k_{x}^2\sin^2\alpha+k_{x}k_z\sin2\alpha+k_z^2\cos^2\alpha\right)}{k_{\rho}^2+k_z^2}-1\right],
 \end{equation}
 and integrating over $dz$, we get an effective 2D NLGPE: 

\begin{equation}
\begin{split}
i\hbar\frac{\partial}{\partial t}\psi(x,y,t)=\left[-\frac{\hbar^2}{2m}\nabla^2_{x,y}+\frac{g}{\sqrt{2\pi}l_z}|\psi(x,y,t)|^2+\right .\\
\left. \frac{2 g_d}{3l_z}\int \frac{dk_xdk_y}{(2\pi)^2}e^{i(k_xx+k_yy)}f(k_x,k_y)\ \tilde n(k_x,k_y)\right]\psi(x,y,t),
\label{2dgpe}
\end{split}
\end{equation}
with $\tilde n(k_x,k_y)$ Fourier transform of $|\psi(x,y)|^2$ and
\begin{equation}
\begin{split}
f(k,\theta_k)=\sqrt{2\pi}\left(3\cos^2\alpha-1\right)+3\pi \ e^{k^2/2}k \ {\rm erfc}\left(\frac{k}{\sqrt{2}}\right)\\
\times\left(\sin^2\alpha\cos^2\theta_k-\cos^2\alpha\right),
\end{split}
\end{equation}
where we use dimensionless polar coordinates $\big(k\equiv l_z\sqrt{k_x^2+k_y^2}$ and $\theta_k\big)$, and $\mathrm{erfc}(x)$ is the complimentary error function. Note that for $\alpha=0$ and $\alpha=\pi/2$ Eq. \eqref{2dgpe} reduces to the cases discussed in Refs. \cite{sol-luis} and \cite{nath-pho}, respectively for the realization of isotropic and anisotropic solitons.

%%%%%%%%%%%%

%%%
 \begin{figure}[t]
\centering
\includegraphics[width= 1\columnwidth]{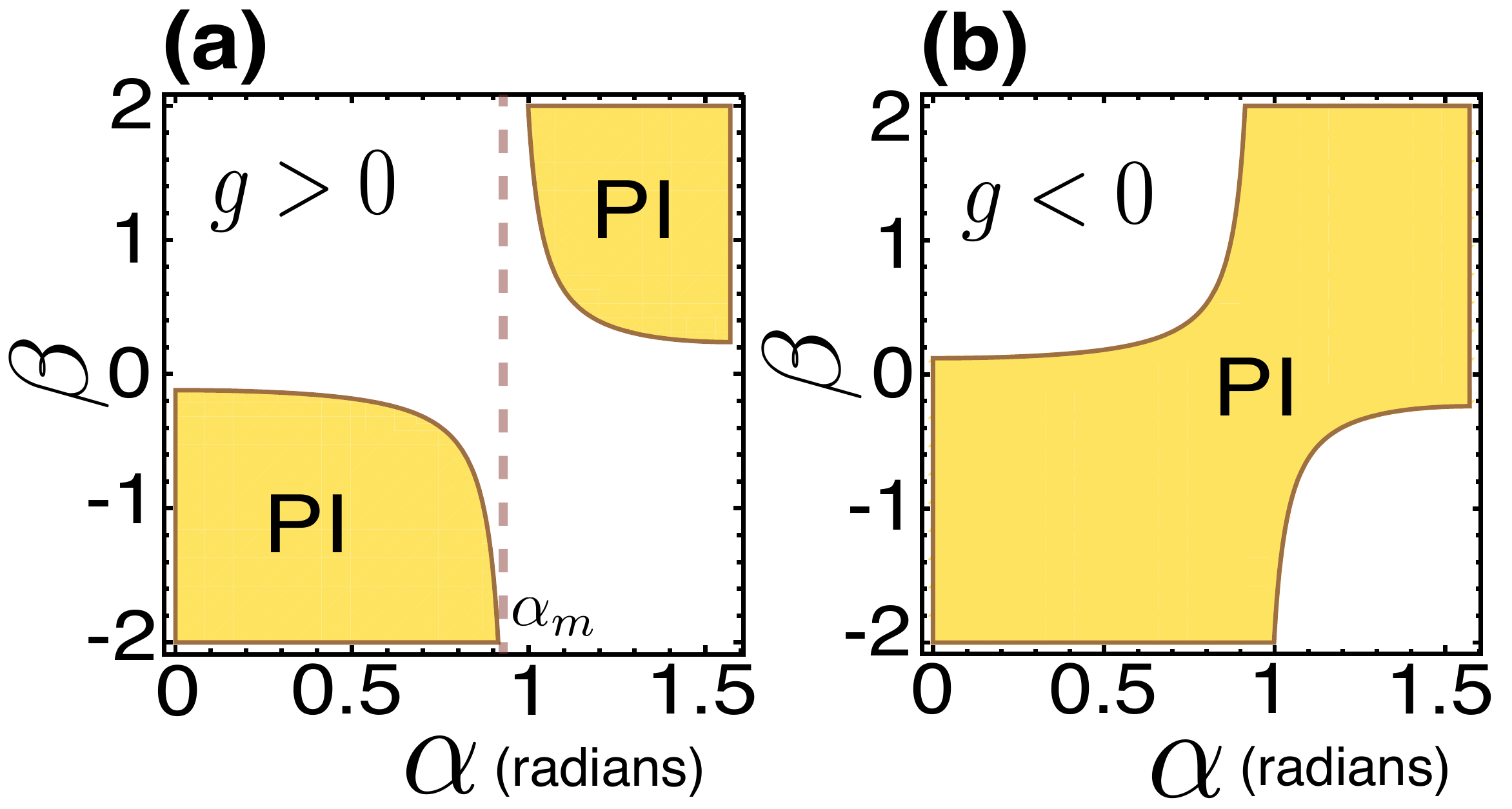}
\caption{\small{The PI regions (shaded) of a 2D dipolar BEC as a function of $\beta$  and the tilting angle $\alpha$ (in radians) for  (a) $g>0$  and (b) $g<0$. In the former case post PI dynamics is characterized by the formation of 2D solitons, while in the latter case the BEC becomes unstable against local collapses. The dashed line in Fig. \ref{fig:ab}(a) indicates the magic angle $\alpha_m$ across which the characteristics of DDI should be reversed in order to stabilize the 2D bright solitons.}}
\label{fig:ab} 
\end{figure}
%%%% 

\section{Phonon Instability}
\label{be-pi}
In this section, we calculate the low energy spectra (Bogoliubov excitations) of a 2D homogeneous dipolar BEC with tilted dipoles. The homogeneous solution of Eq. \eqref{2dgpe} is $\psi(x,y,t)=\sqrt{n_{2D}}\exp[-i\mu_{2D}t/\hbar]$, with $n_{2D}$ the homogeneous 2D density. The 2D chemical potential is obtained as
\begin{equation}
\mu_{2D}=\frac{gn_{2D}}{\sqrt{2\pi}l_z}\left[1+\frac{4\pi}{3}\beta\left(3\cos^2\alpha-1\right)\right],
\end{equation}
where $\beta=g_d/|g|$ determines the ratio between the DDI strength and that of the contact interactions. The elementary excitations of the homogeneous solution are of the form $\delta\psi_k(x,y,t)=u_ke^{-i({\bf k}\cdot{\pmb \rho}-\omega_{\bf k} t)}+v_k^*e^{i({\bf k}\cdot{\pmb\rho}-\omega_{\bf k}^* t)}$ with the dispersion
\begin{equation}
\epsilon_{\bf k}=\hbar\omega_{\bf k}=\sqrt{E_k\left\{E_k+\frac{2gn_{2D}}{\sqrt{2\pi}l_z}\left[1+{\frac{2\sqrt{2\pi}}{3}}\beta f(k)\right]\right\}},
\end{equation}
where $E_k=\hbar^2\left(k_x^2+k_y^2\right)/2m$. The phonon modes, $\epsilon({\bf k}\to 0)\propto \sqrt{\mu_{2D}}k_{\rho}$ when $\mu_{2D}<0$, which provides the PI condition:
\begin{equation}
 g < -\frac{4\pi g_d}{3} \left(3\cos^2\alpha-1\right).
 \label{pic}
\end{equation}

For $g<0$ PI always leads to local collapses, resembling the situation in BECs with short-range attractive interactions \cite{coll_corn_01,coll_hul_00} and in 3D homogeneous dipolar BECs \cite{dip_rvw_09,dip_san_00}. We hence focus below on BEC with repulsive contact interactions, $g>0$, where PI 
arises due to the attractive part of the anisotropic DDI. As shown in Ref.~ \cite{nath-pho}, for $\alpha=0$ and $\alpha=\pi/2$ the post PI dynamics is characterized by the formation of a transient  gas of bright solitons.  
The solitons then eventually undergo inelastic collisions, fusing together to form larger ones. The larger solitons may survive against collapse if the BEC density is low enough to hold the 2D criteria, 
$|\mu_{2D}|\ll\hbar\omega_z$. 
 
The PI regions as a function of $\beta$ and $\alpha$ with $g>0$ (Fig. \ref{fig:ab}a) provide the first estimation for the stability regions of 2D bright solitons. In Fig. \ref{fig:ab}(a), there are 
two different regions in the $\alpha-\beta$ plane satisfying the PI criteria, separated at the \emph{magic angle} $\alpha_m=54.7$ degrees (or 0.95 radians). For $\alpha<\alpha_m$ the PI requires $g_d<0$ which demands inverting the anisotropic character of DDI via rotating fields. This makes the direct cross-over between the two PI regions impossible by continuously varying $\alpha$. Note that, in both regions, a sufficiently large $|\beta|$ may lead to local collapses. This introduces an upper cut-off for $|\beta|$ for the creation of stable solitons, which will be estimated in Sec.~\ref{2dsol} using variational calculations. 
Note from Fig.~\ref{fig:ab}(a) that a dipolar BEC may be driven into PI just by tilting the dipoles from an initially stable configuration, without changing the interaction parameters or trap geometry. 
As it can be easily accomplished by changing the orientation of the externally applied field, we propose this as an alternative simple method to generate 2D bright solitons in dipolar BECs, combined with an adiabatic approach. This is numerically investigated in Sec. \ref{solcr}.

%%%%%%

\section{Two dimensional bright solitons with tilted dipoles}
\label{2dsol}
In this section, we analyze the stability of 2D solitons in a dipolar BEC with tilted dipoles, using a 3D variational solution and 
numerical NLGPE solutions. 

%%%%%%%
\subsection{Gaussian ansatz}
\label{gauss}
We consider the following Gaussian ansatz:
\begin{equation}
\Psi_0({\bf r})=\frac{1}{\pi^{3/4}l_z^{3/2}\sqrt{L_xL_yL_z}}\exp\left[-\frac{1}{2l_z^2}\left(\frac{x^2}{L_x^2}+\frac{y^2}{L_y^2}+\frac{z^2}{L_z^2}\right)\right].
\label{gau}
\end{equation} 
The dimensionless variational parameters: $L_x$, $L_y$ and $L_z$ provide the Gaussian widths along $x$, $y$ and $z$. 
Introducing this ansatz in the energy functional
\begin{equation}
\begin{split}
E=\int d^3r\left[\frac{\hbar^2}{2m}|\nabla\Psi_0(r)|^2+V_t(r)|\Psi_0(r)|^2+\frac{g}{2}|\Psi_0(r)|^4+\right. \nonumber \\
\left.\frac{1}{2}\int d^3r^{\prime}V_d(r-r^{\prime})|\Psi_0(r)|^2|\Psi_0(r^{\prime})|^2\right],
\end{split}
\end{equation}
we obtain
\begin{widetext}
\begin{equation}
\begin{split}
\frac{E}{\hbar\omega_z}=\frac{1}{4L_x^2}+\frac{1}{4L_y^2}+\frac{1}{4L_z^2}+\frac{L_z^2}{4}+\frac{\tilde g}{4\pi L_xL_yL_z}+\frac{\tilde{g}_d}{3L_z}\left[\frac{3\sin^2\alpha}{L_x^2-L_y^2}\left(\sqrt{\frac{L_y^2-L_z^2}{L_x^2-L_z^2}}-\frac{L_y}{L_x}\right)+\frac{3\cos^2\alpha}{\sqrt{\left(L_x^2-L_z^2\right)\left(L_y^2-L_z^2\right)}}-\frac{1}{L_xL_y}\right] \\
-\frac{2\tilde g_d}{\pi}\int_0^{\pi/2}d\chi\frac{\cos^2\chi\sin^2\alpha-\cos^2\alpha}{\left[L_z^2-\left(L_x^2\cos^2\chi+L_y^2\sin^2\chi\right)\right]^{3/2}} \ \  {\rm arctanh}\left(\frac{\sqrt{L_z^2-\left(L_x^2\cos^2\chi+L_y^2\sin^2\chi\right)}}{L_z}\right),
\end{split}
\label{E3d}
\end{equation}
\end{widetext}
where $\tilde g=g/\sqrt{2\pi}\hbar\omega_zl_z^3$ and $\tilde g_d=g_d/\sqrt{2\pi}\hbar\omega_zl_z^3$. The minimum of $E(L_x, L_y, L_z)$ provides the equilibrium widths $\{w_x^0, w_y^0,w_z^0\}$ of the soliton. The absence of this minimum results in  two distinct types of instability. If the repulsive part of the interactions dominates, the soliton expands without limits on the $xy$ plane ($w_{x,y}^0\to\infty$), whereas dominating attractive interactions lead to collapse ($w_{x,y,z}^0\to 0$). 

%%%%%%%%
\subsection{Stability analysis and properties of 2D solitons}
\label{staban}
For $\alpha=0$, we recover the isotropic scenario of Ref.~\cite{sol-luis}, where stable isotropic solitons demand
\begin{equation}
\frac{2\tilde g_d}{3\sqrt{2\pi}} < 1+\frac{\tilde g}{(2\pi)^{3/2}} < \frac{-4\tilde g_d}{3\sqrt{2\pi}},
\end{equation}
which requires $\tilde g_d<0$, i.e. the inversion of the DDI. For large values of $g$, the above criteria reduces to $\beta<-3/8\pi$. 
Any $\alpha>0$ breaks polar symmetry, and hence $w_x^0\neq w_y^0$. 
We quantify the 2D soliton anisotropy with the aspect ratio $\gamma=w_{i}^0/w_{j}^0$, where $\{i, j\} \in \{x,y\}$ with $w_j^0>w_i^0$ such that $\gamma\leq 1$.

\begin{figure}[ht]
\centering
\includegraphics[width= 0.9\columnwidth]{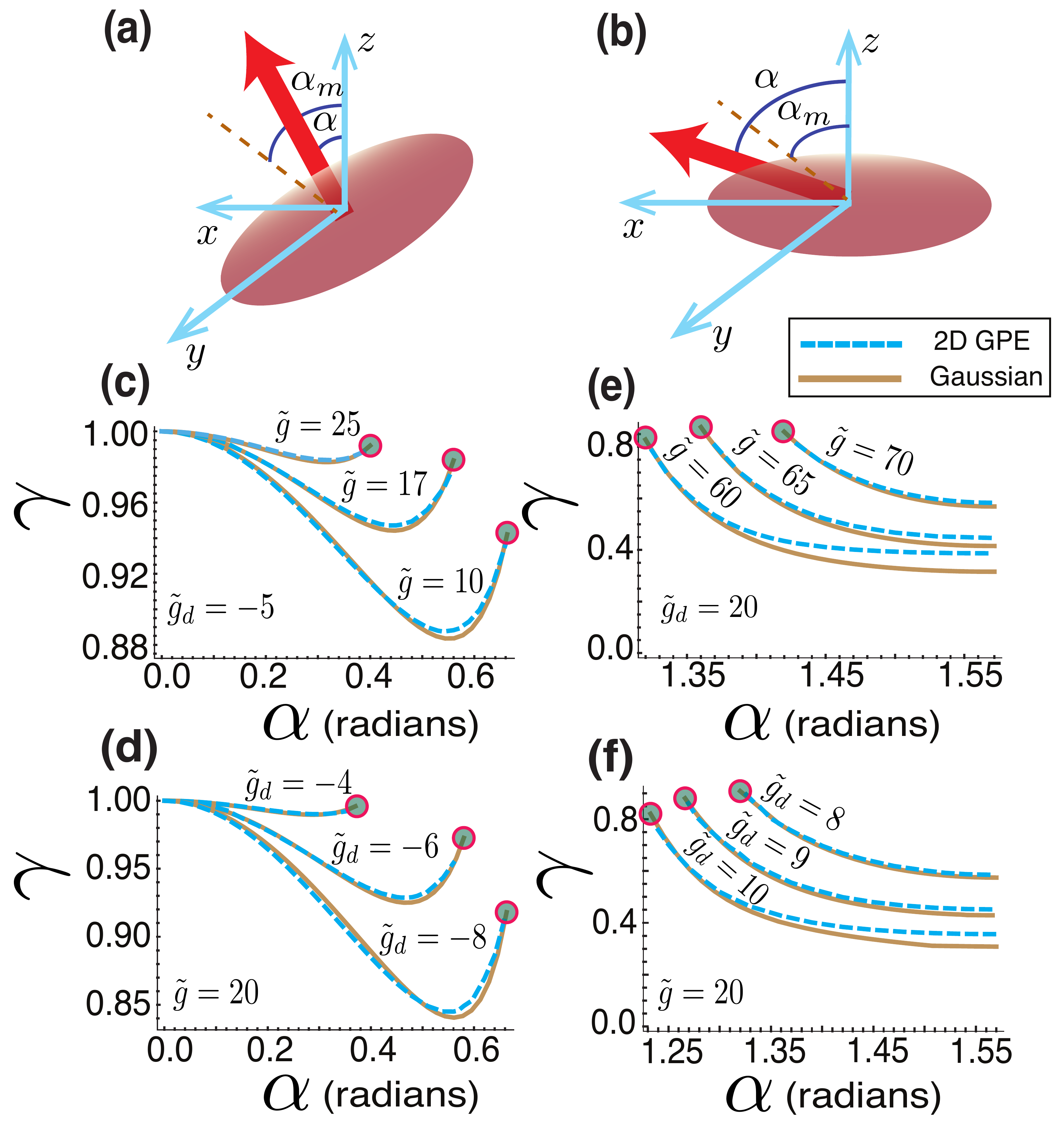}
\caption{\small{The figures (a) and (b) show the equilibrium configurations of the soliton for (i) $\beta<0$ $(0<\alpha<\alpha_m)$ and (ii) $\beta>0$ $(\alpha_m<\alpha\leq\pi/2)$. The thick (red) arrow shows the orientation of the dipoles in the BEC. For the case (i) the soliton is more elongated along the $y$ axis and hence the aspect ratio is taken as $\gamma=w_x^0/w_y^0$. In figures (c) and (d) $\gamma$ as a function of $\alpha$ is shown for the case (i). The value of $\alpha$ at which each curves terminates ($\alpha=\alpha_e$), shown by filled circles, gives the critical angle for the expansion instability. For case (ii) the soliton is more elongated along the $x$ axis, and hence we define $\gamma=w_y^0/w_x^0$, the corresponding plots as a function of $\alpha$ are shown in (e) and (f). In this case for $\alpha<\alpha_e$ the solitons are unstable against expansion.}}
\label{fig:lxlya} 
\end{figure}

For $\beta<0$, increasing $\alpha$ from zero towards $\alpha_m$ leads to soliton elongation along $y$, $w_y^0>w_x^0$, since the DDI becomes more attractive along $y$ than along $x$, as schematically shown in Fig. \ref{fig:lxlya}a. In addition, both $w_x^0$ and $w_y^0$ increase monotonously with $\alpha$ since the overall attractive interaction is reduced. As shown in Fig.~\ref{fig:lxlya}c, $\gamma$ shows a non-monotonous character with a minimal value. Near the expansion instability, which happens at a $\tilde g$- and $\tilde g_d$-dependent angle $\alpha_e$ ($<\alpha_m$ and shown by filled circles in  Fig. \ref{fig:lxlya}c-f), the soliton anisotropy diminishes. The latter occurs because close to $\alpha_e$ both $w_{x,y}^0$ are very large, the interaction energy becomes hence very small, and as a result the anisotropy diminishes.
For $\beta>0$, for $\alpha=\pi/2$ the solitons are maximally anisotropic with $w_x^0>w_y^0$ for any value of $\tilde g$ and $\tilde g_d$ (see Fig. \ref{fig:lxlya}b). 
As expected, when $\alpha$ decreases from $\pi/2$, the anisotropy decreases monotonously, until instability against expansion occurs at a critical angle.
Figures \ref{fig:lxlya}c (d) and \ref{fig:lxlya}e (f) show the results for different $\tilde g$ ($\tilde g_d$) values with a fixed $\tilde g_d$ ($\tilde g$). In both figures we show the results for $\gamma$ obtained from 
the variation ansatz discussed above and from the numerical simulation of the 2D NLGPE, which are excellent agreement in basically for all cases. As expected, the anisotropy increases with growing $|\tilde g_d/\tilde g|$.
The fact that the values of $\tilde g$ and $\tilde g_d$ for with the 2D soliton is stable depend on the tilting angle $\alpha$, allows for the observation of solitons in parameter regimes 
in which solitons are unstable either for $\alpha=0$  \cite{sol-luis} or $\alpha=\pi/2$  \cite{sol-aniso}, hence easing the experimental realization of 2D solitons.

\subsection{Soliton gas formation after phonon instability}

\begin{figure}[hbt]
\centering
\includegraphics[width= 1.\columnwidth]{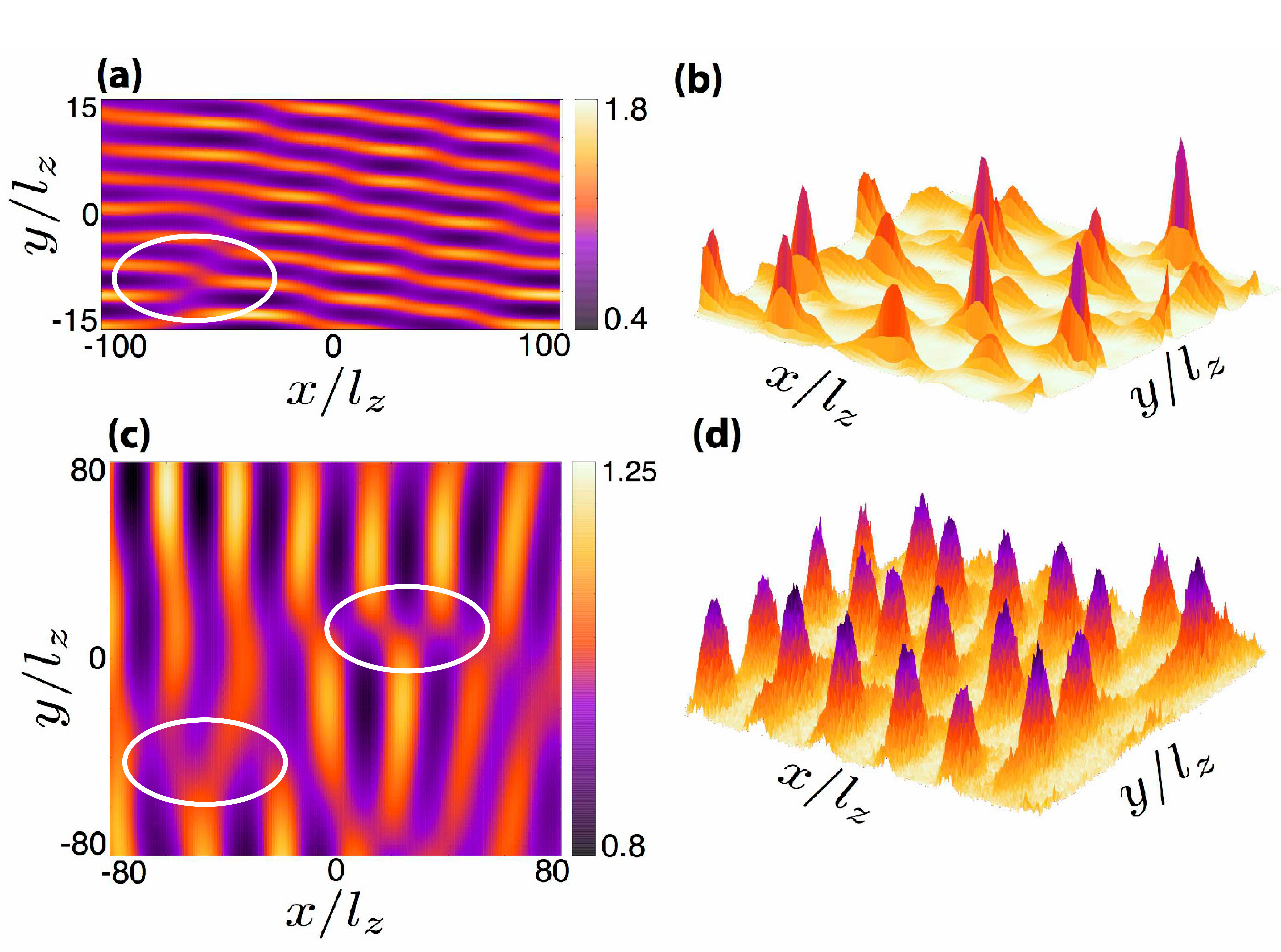}
\caption{\small{The snapshots of the post-PI dynamics in a 2D dipolar homogeneous condensate. The figures (a) and (b) are for $\tilde g=12$, $\beta=0.28$ and $\alpha=1.35$ radians at times $t=13.5/\omega_z$ and $t=27/\omega_z$ respectively. Similarly, the figures (c) and (d) are for $\tilde g=40$, $\beta=-0.29$ and $\alpha=0.6$ radians at times $t=31/\omega_z$ and $t=90/\omega_z$ respectively. Figures (a) and (c) are the transient stripes patterns with the dislocation defects shown in ellipses. Figures (b) and (d) are the temporarily arranged, unstable, ordered state of the soliton gas, which eventually merge each other during collisions.}}
\label{fig:pid} 
\end{figure}

As mentioned above, PI in 2D dipolar BECs may be followed by the formation of a transient gas of bright solitons, instead of the collapse characteristic of 
short-range interacting BECs \cite{nath-pho}. The post-instability evolution and the dynamics of the emergent soliton gas depend crucially on the tilting angle $\alpha$. 
Figs. \ref{fig:pid}(a) and  \ref{fig:pid}(b) show the post-instability density patterns for $\tilde g=12$, $\beta=0.28$ and $\alpha=1.35$ radians at times $t=13.5/\omega_z$ and $t=27/\omega_z$, respectively.
Note that the the first stages of the post-instability dynamics are characterized by the formation of a transient stripe pattern that present dislocation defects \cite{pat_form_bk}. At these dislocations 
two stripes merge into one. Note that for the case considered the DDI is more attractive along the $x$-axis, which results in stripes almost parallel to the $x$ direction.  The initial formation of stripes is characteristic of in-plane anisotropy resulting from $\alpha>0$. Stripes may be observed as well for $\beta<0$, as illustrated in Figs. \ref{fig:pid}(c) and (d) for $\beta=-0.29$ and $\alpha=0.6$ radians at times$t=31/\omega_z$ and $t=90/\omega_z$, respectively.  In this case, the DDI is more attractive along the $y$ direction, resulting in stripes along the same direction.
As shown in Fig. \ref{fig:pid}(b) and (d), the density stripes eventually break down into anisotropic solitons with major axis along the direction in which the DDI is more attractive, resulting in an unstable ordered state of solitons. The solitons eventually attract each other, fuse together and may remain stable depending on its condensate density. We have observed that the dynamics of the soliton gas inherently slowed down for $\alpha\neq 0$ compared to the isotropic case ($\alpha=0$). This is because, in the latter case the attractive forces between the solitons are rotationally invariant, whereas in the former case it is dominant along one particular direction due to the anisotropy of DDI, which restricts the motion in other directions.

%%%%%%%%%%%%%%%%%
\subsection{Variational Calculations : Low-lying excitations}
\label{modes}

\begin{figure}[t]
\centering
\includegraphics[width= 1.\columnwidth]{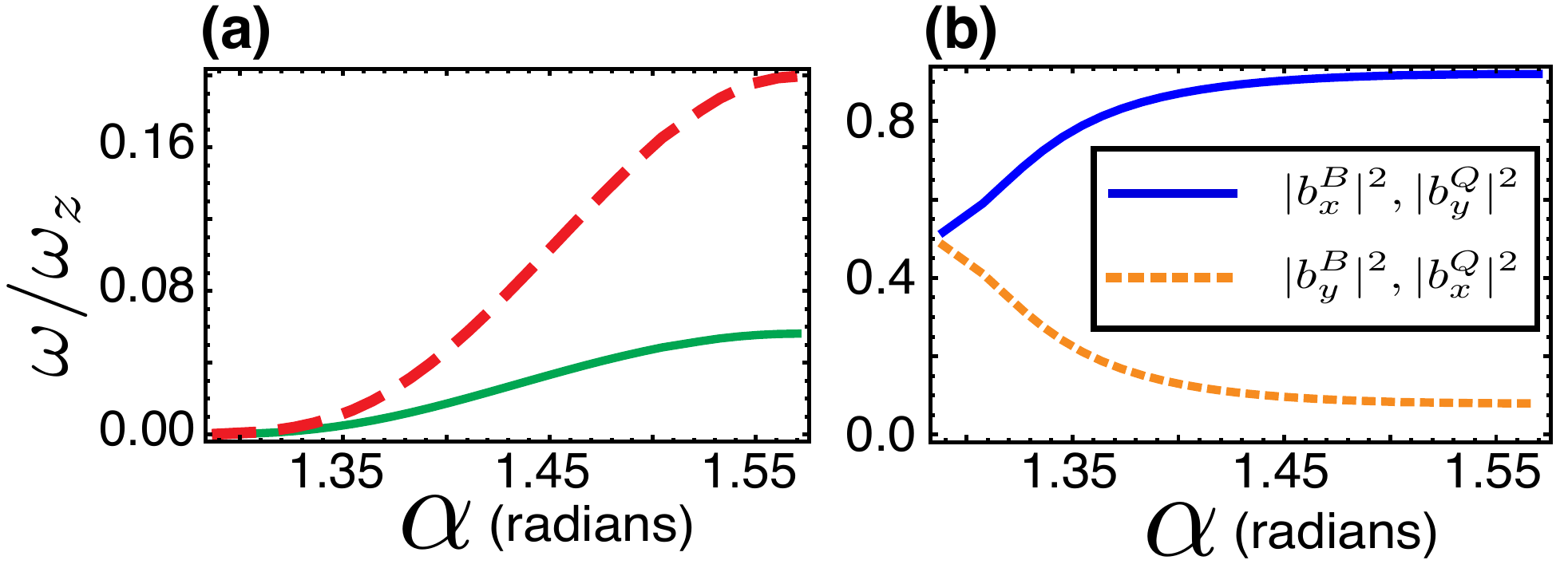}
\caption{\small{ The eigen frequencies (a) and the square of the components of the eigen-vectors (b) of two low-lying $xy$ modes for $\beta=0.36$ as a function of $\alpha$. For $\alpha=\pi/2$, the soliton is maximally anisotropic and the modes exhibit a pure $x$ (solid lines) and $y$ (dashed lines) character. As $\alpha$ reduces, the solitons become less anisotropic and near the expansion instability the modes turn into breathing and quadrupole like ones, akin to that of the isotropic case.}}
\label{fig:atmxy} 
\end{figure}

At this point we examine the lowest-lying modes of 2D dipolar solitons using a variational method \cite{var_zoll_96,var_you_01}, where we use a time-dependent Gaussian as the trial wave function: 
\begin{eqnarray}
\psi(x,y,z,t) &=& A(t)\prod\limits_{\eta=x,y,z} e^{-\frac{(\eta-\eta_0(t))^2}{2w_\eta^2(t)}} e^{i\eta\alpha_\eta(t)} e^{i\eta^2\beta_\eta(t)},
\end{eqnarray}
where $\eta_0$, $w_\eta$, $\alpha_\eta$ and $\beta_\eta$ are the time dependent variational parameters and the normalization constant $A(t)=\pi^{-3/4}/\sqrt{w_xw_yw_z}$. We consider the soliton is static in the $xy$ plane and assume $x_0(t)=0$ and $y_0(t)=0$. The above ansatz is then introduced in the Lagrangian density of a dipolar BEC:
\begin{widetext}
\begin{eqnarray}
\mathcal L&=& \frac{i}{2}\hbar\left( \psi\frac{\partial{\psi^*}}{\partial{t}}-\psi^*\frac{\partial{\psi}}{\partial{t}}\right)+\frac{\hbar^2}{2m}|\nabla\psi(r,t)|^2 + V_t(r)|\psi(r,t)|^2+\frac{g}{2}|\psi(r,t)|^2+\frac{1}{2}|\psi(r,t)|^2 \int dr' V_d(r-r')|\psi(r',t)|^2\nonumber.
\label{lad}
\end{eqnarray}
\end{widetext}
The  Lagrangian is then obtained by integrating over the whole space i.e., $L=\int d^3r\mathcal L$. We then obtain the corresponding Euler-Lagrange equations of motion 
for the time-dependent variational parameters, see App. \ref{lang}. Note that the centre of mass motion (along the $z$ axis) is decoupled from the internal dynamics of the soliton. Below we analyze in detail the two in-plane ($xy$-) modes of the condensate, which lie lowest in the excitation spectrum. 
For $\alpha=0$  (isotropic case), those are the breathing and quadrupole modes \cite{sol-luis}. When $\alpha\neq 0$, the character of the in-plane modes change, and in particular as the anisotropy of the soliton increases the modes in the $xy$ plane decouple into pure $x$ and $y$ modes. In order to visualize it, we plot the square of the components of the eigen-vectors ${\bf b}=b_x \hat x + b_y\hat y$ of the two planar modes for a particular case with $\beta=0.36$, see Fig. \ref{fig:atmxy}(b) and the corresponding eigen-frequencies ($\omega$) are shown in Fig. \ref{fig:atmxy}(a). For $\beta=0.36$ and $\alpha=\pi/2$, the condensate is more elongated along the $x$-direction and hence the lowest mode (solid lines in Fig. \ref{fig:atmxy}) accounts mostly the oscillation of condensate width along the $x$-axis. When $\alpha$ reduces or dipoles tilting out of the $xy$ plane, the anisotropy of the soliton decreases and the modes become more breathing-like (the lowest one) and quadru
 pole-like modes of the isotropic case, Fig. \ref{fig:atmxy}(b). Near the expansion instability they almost turn into breathing and quadrupole modes. The eigen-frequencies of the low lying modes are also crucial for the adiabatic control or preparation of the dipolar soliton, as discussed in Sec. \ref{solcr}, where we have taken $\alpha\equiv \alpha(t)$. In particular, the frequency of the lowest mode determines the rate at which $\alpha$ should be varied in time in order for the system to remain in the instantaneous ground state of the system.

%%%%%%%%%%%%%%%%%%%%%%%%%%%%
\subsection{Experimentally Relevant atomic systems}
\label{exp}
\begin{figure*}[hbt]
\centering
\includegraphics[width= .95\textwidth]{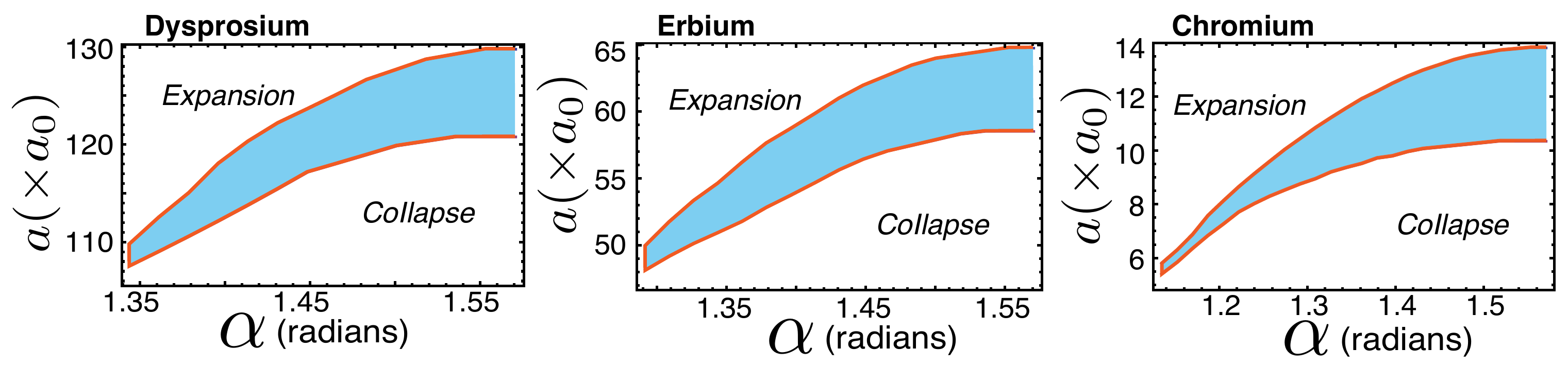}
\caption{\small{The stability regions (shaded) for the anisotropic solitons as a function of tilting angle $\alpha$ and the $s$-wave scattering length $a$ (in units of Bohr radius $a_0$) for Dysprosium, Erbium and Chromium condensates, with $\omega_z=2\pi\times 1$kHz and $N=10000$. The dipolar interaction is calculated with the intrinsic magnetic moments of each of the atoms. For large values of $a$ the soliton become unstable against expansion and for lower values it undergoes the collapse instability.}}
\label{fig:atms} 
\end{figure*}

Now we discuss the stability regions of the soliton in the state-of-the-art experimentally realized  dipolar condensates of Cr, Dy and Er. In particular, we focus on the anisotropic solitons with $\beta>0$, as it is experimentally less challenging. We take $\omega_z=2\pi\times 1$kHz and $N=10000$, resulting in an effective dipolar interaction strengths of $\tilde g_d=$ 22 (Cr), 170 (Er), 334 (Dy) for different atoms. The resulting phase diagrams obtained from Gaussian-energy calculations are shown in Fig. \ref{fig:atms} as a function of the $s$-wave scattering length $a$ and the tilting angle $\alpha$. These phase diagrams are of considerable interest if one would like to tune the system from soliton unstable to soliton stable region either by tilting the dipoles orientation or tuning $a$ using the Feshbach resonance. Below we demonstrate the adiabatic preparation of the soliton in a Cr BEC, using experimentally realistic parameters, by tuning the tilting angle in time in which the dipolar BEC undergoes the transition from externally confined BEC to a self trapped one.

%%%%%%%%%%%%%%%
\section{Adiabatic preparation of an anisotropic bright soliton}
\label{solcr}
\begin{figure}[t]
\centering
\includegraphics[width= 1.\columnwidth]{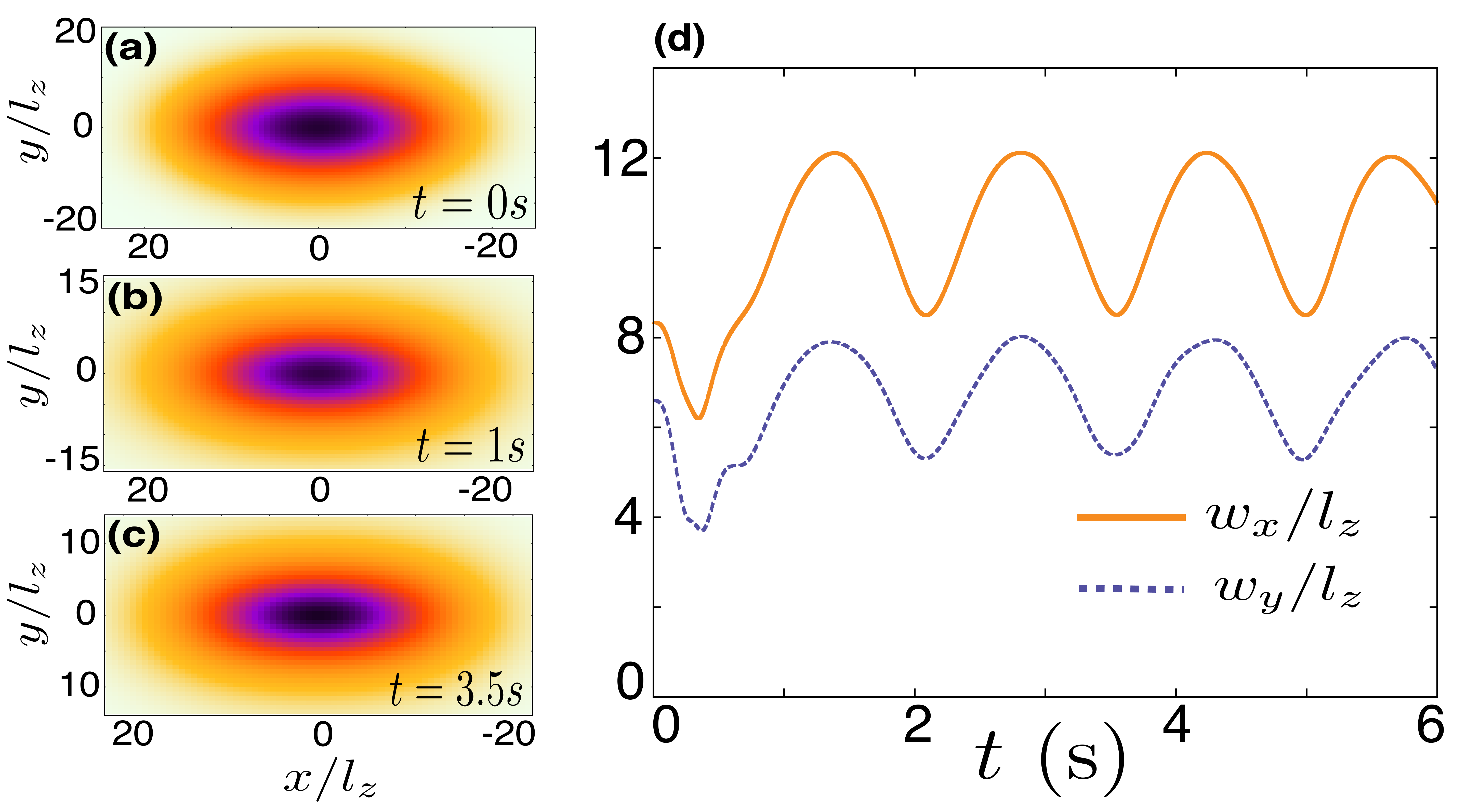}
\caption{\small{The numerical results of 2D GPE for the preparation of an anisotropic soliton starting from a weakly confined ($\omega_{x,y}=2\pi \times 10$Hz) Cr BEC with $N=6000$, $a=9 a_0$ and $\alpha=1.22$ radians. (a) The initial condensate density, (b) after 1s and (c) after 3.5s. The tilting angle $\alpha(t)$ is varied linearly from $\alpha_i=1.22$  to $\alpha_f=1.32$ radians in 0.3s. Then, the trap in the $xy$ plane is removed in 0.5s.  (d) The widths $w_x$ and $w_y$ of the soliton along $x$ and $y$ directions respectively as a function of time. The periodic oscillation of the widths for  $t> 0.8$s indicates the self-trapping of the condensate.}}
\label{fig:solt} 
\end{figure}

The dependence of the BEC stability and of the soliton properties on the tilting angle $\alpha$ opens interesting perspectives for the preparation of 2D solitons by 
modifying in real time the tilting angle. In contrast to the previous sections we consider at this point the presence of an external shallow trap on the $xy$ plane with frequencies $\omega_{x,y}=2\pi\times 10$ Hz. 
We consider a Cr-BEC of $N=6000$ atoms with $\omega_z=2\pi\times 1$ kHz, initially prepared with a tilting angle $\alpha_i=1.22$ radians and $a=9 a_0$, such that under these conditions there is no self-trapping (i.e. no 2D soliton), and the only confinement is provided by the 
$xy$ trap. We may then keep the interaction parameters intact, and slowly tune $\alpha$ to a final value ($\alpha_f=1.32$ radians) within the soliton-stability region. The trap on the $xy$ plane may be then 
removed. The latter should be done slow-enough such that one avoids the significant creation of excitations in the condensate that may affect the self-trapping. The numerical results using the real-time evolution of the 2D NLGPE are shown in Fig. \ref{fig:solt}, where we monitor the widths of the condensate in the $xy$ plane as function of time. 
The self-trapping of the condensate in the $xy$ plane is evident from the periodic oscillation of the condensate widths, after the removal of the harmonic confinement. 
Note that the created soliton presents a slight breathing motion on the xy plane and the numerical results using the 2DGPE for the condensate density at various instants are shown in Fig. \ref{fig:solt} a-c. 

We can find different regimes of adiabaticity for the creation of bright soliton in dipolar BEC based on the different time scales in the system \cite{adia_02}. Here, we employ the condition $T_{\omega} \ll T \ll T_{\mu}$, where $T_{\omega}=\max\{1/\omega_B(\alpha(t))\}$, the equivalent of quantum mechanical linear adiabatic time scale, determined by the inverse of the lowest collective (breathing) mode $\omega_B$ and the nonlinear time scale  $T_{\mu}=\max\{1/|\mu_{2D}(\alpha(t))|\}$ depends on the instantaneous chemical potential $\mu_{2D}(t)$. Note that in our case we vary the tilting angle from $\alpha_i$ to $\alpha_f$  in which the chemical potential changes from a positive to negative value continuously through $\mu_{2D}=0$ ($T_{\mu}=\infty$). Thus the adiabatic criteria simply becomes  $T_{\omega} \ll T$, which guarantee us that we do not significantly populate the excitations during the period $T$ \cite{ad_nl_07}. In the particular case considered here, $T_{\omega}\sim 0.05$s. The soliton preparation involves two steps, first in which we vary $\alpha$ and the second involves the removal of $xy$ confinement, both done linearly in time. We varied $\alpha$ in 0.3s from $\alpha_i$ to $\alpha_f$ and the trap is then removed in a duration of 0.5s.

%%%%%%%%%%%%%

%%%%%%%%%%%%%%%%%

%%%%%%%%%%%%
\section{Conclusions}
\label{con}
In conclusion, we studied the physics of 2D  bright solitons in dipolar condensates as a function of the orientation of the dipoles with respect to  the soliton plane. As we showed, the tilting angle may enhance the experimental possibilities  to observe the self-trapped matter waves as well as provide a probe to tune the anisotropy of the soliton. In addition, it can drive the dipolar BEC into PI without altering its interaction parameters or trap geometry.  
In 2D, the post PI dynamics is always characterized by a transient stripe formation and eventual formation of an unstable soliton gas. Finally we have demonstrated how to prepare the 2D soliton adiabatically by tuning the tilting angle in the case of a Cr BEC.

%%%%%%
\section{Acknowledgments}
R. N and P. P acknowledge funding by the Indo-French Centre for the Promotion of Advanced Research - CEFIPRA. L.S. thanks the support of the Cluster QUEST, and  
the Deutsche Forschungsgemeinschaft (RTG 1729). M. R. acknowledges the funding from DST India through INSPIRE scholarship. Also, P. P.  acknowledges financial support from Conseil R\'{e}gional d'Ile-de-France under DIM Nano-K / IFRAF, CNRS, and from Minist\`{e}re de l'Enseignement Sup\'{e}rieur et de la Recherche within CPER Contract.
%%%%%%%%%%%%
%% APPENDIX 
%%%%%%%%%%%%
\appendix
\section{Lagrangian for a dipolar condensate}
\label{lang}
The Lagrangian density for a dipolar condensate is given in Eq. \ref{lad}, and using the Gaussian time-dependent trial function we obtain the Lagrangian $L=\int d^3r\mathcal L$:
\begin{eqnarray}
L &=& \sum_{\eta=x,y,z}\left[\hbar\frac{\dot{\beta}_\eta w_\eta^2}{2}+\frac{\hbar^2}{2m} \left( \frac{1}{2w_\eta^2}+ \alpha_\eta^2 + 2\beta_\eta^2 w_\eta^2 \right)\right] +\frac{1}{2}m\omega_z^2 \frac{w_z^2}{2}\nonumber \\
&&+ \frac{g}{\sqrt{2\pi}}\frac{1}{4\pi w_x w_yw_z} + V(w_\eta), \nonumber \\
\label{lagrangian}
\end{eqnarray}
where 
\begin{equation}
 V(w_\eta)= \frac{1}{2}\frac{1}{\left(2\pi\right)^3} \int d^3k \tilde V_d({\bf k}) \prod\limits_{\eta=x,y,z}e^{-\frac{k_\eta^2 w_\eta^2}{2}}
\end{equation}
with $\tilde V_d({\bf k})$ the Fourier transform the dipole-dipole potential. Then, the equations of motion for the condensate widths are obtained as
\begin{eqnarray}
m\ddot w_x &=& \frac{\hbar^2}{mw_x^3} + \frac{g}{(2\pi)^{3/2} w_x^2w_yw_z} - 2\frac{\partial{V}}{\partial{w_x}} \\
m\ddot w_y &=& \frac{\hbar^2}{mw_y^3} + \frac{g}{(2\pi)^{3/2} w_xw_y^2w_z} - 2\frac{\partial{V}}{\partial{w_y}} \\
m\ddot w_z &=& \frac{\hbar^2}{mw_z^3} + \frac{g}{(2\pi)^{3/2} w_xw_yw_z^2} -m\omega_z^2z^2- 2\frac{\partial{V}}{\partial{w_z}}.
\end{eqnarray}
The above equations describe the motion of a particle with coordinates $w_\eta$ in an effective potential
\begin{eqnarray}
U(W_\eta) &=&  \frac{\hbar^2}{2m}\sum_\eta\frac{1}{w_\eta^2}+ \frac{1}{2}m\omega_z^2w_z^2+ \frac{g}{(2\pi)^{3/2} w_xw_yw_z} + V(w_\eta).\nonumber \\
\end{eqnarray}
Once the equilibrium widths of the condensate are obtained by minimizing the effective potential (or equivalently from the Gaussian energy calculations in Section. \ref{gauss}), the low lying excitations  are obtained  by diagonalizing the Hessian matrix of $U$. Also, note that the centre of mass motion of the soliton along the $z$ axis is de-coupled from the internal dynamics and is governed by the equation
\begin{equation}
\ddot z_0=-\omega_z^2z_0.
\end{equation}
\bibliographystyle{apsrev}
\bibliography{liball.bib}

\end{document}